\documentclass[%
 reprint,%
 amssymb, amsmath,%
 aps,cha,%
]{revtex4-1}
\usepackage{graphicx}% Include figure files
\usepackage{epstopdf}
\usepackage{bm}%
\usepackage{xcolor}

\expandafter\ifx\csname package@font\endcsname\relax\else
 \expandafter\expandafter
 \expandafter\usepackage
 \expandafter\expandafter
 \expandafter{\csname package@font\endcsname}%
\fi
\begin{document}

%\preprint{}

\title{
%Raman pulse atom interferometry on the rubidium blue transitions\\
%oppure\\
Atom Interferometry with the Rb Blue Transitions
%oppure\\
%Raman-Pulse Atom Interferometry with the Rubidium Blue Transitions
} %Title of paper

\author{L. Salvi$^1$}

\author{L. Cacciapuoti$^2$}

\author{G. M. Tino$^1$}
\thanks{Also at: CNR-INO, Sesto Fiorentino, Italy}

\author{G. Rosi$^1$}
\thanks{corresponding author}

\affiliation{
$^1$ Dipartimento di Fisica e Astronomia and LENS, Universit\`{a} di Firenze,\\ INFN Sezione di Firenze, via Sansone 1, I-50019 Sesto Fiorentino (FI), Italy\\
$^2$ European Space Agency, Keplerlaan 1, 2201 AZ Noordwijk, Netherlands
}

%\author{L. Salvi}
% \affiliation{
%Dipartimento di Fisica e Astronomia and LENS, Universit\`{a} di Firenze, INFN Sezione di Firenze, via Sansone 1, I-50019 Sesto Fiorentino (FI), Italy
%}
%
%\author{L. Cacciapuoti}
%\affiliation{%
%European Space Agency, Keplerlaan 1, 2201 AZ Noordwijk, Netherlands
%}
%
%\author{G. M. Tino}
%\thanks{Also at: CNR-INO, Sesto Fiorentino, Italy}
% \affiliation{
%Dipartimento di Fisica e Astronomia and LENS, Universit\`{a} di Firenze, INFN Sezione di Firenze, via Sansone 1, I-50019 Sesto Fiorentino (FI), Italy
%}
%
%\author{G. Rosi}
%\thanks{corresponding author}
%\affiliation{
%Dipartimento di Fisica e Astronomia and LENS, Universit\`{a} di Firenze, INFN Sezione di Firenze, via Sansone 1, I-50019 Sesto Fiorentino (FI), Italy
%}

%\date{\today}

\begin{abstract}
We demonstrate a novel scheme for Raman-pulse and Bragg-pulse atom interferometry based on the $5\mathrm{S} - 6\mathrm{P}$ blue transitions of $^{87}$Rb that provides an increase by a factor $\sim 2$ of the interferometer phase due to accelerations with respect to the commonly used infrared transition at 780~nm. A narrow-linewidth laser system generating more than 1~W of light in the 420-422~nm range was developed for this purpose.
Used as a cold-atom gravity gradiometer, our Raman interferometer attains a stability to differential acceleration measurements of $1\times10^{-8}$ $g$ at 1~s and $2\times 10^{-10}$ $g$ after 2000~s of integration time. When operated on first-order Bragg transitions, the interferometer shows a stability of $6\times10^{-8}$~g at 1~s, averaging to $1\times10^{-9}$~g after 2000~s of integration time.  The instrument sensitivity, currently limited by the noise due to spontaneous emission, can be further improved by increasing the laser power and the detuning from the atomic resonance. The present scheme is attractive for high-precision experiments as, in particular, for the determination of the Newtonian gravitational constant.
%\todo[inline]{GMT: modificato un po' l'abstract}
\end{abstract}


\maketitle %\maketitle must follow title, authors, abstract and \pacs

Atom interferometry \cite{tino2014atom}, first demonstrated three decades ago, is now the operating principle of advanced quantum sensors for fundamental physics experiments \cite{Tino2021} and applications \cite{Bongs2019}.
It is used to measure the gravitational acceleration \cite{kasevich1992measurement,peters1999measurement,mueller2008atom,le2008limits,d2019measuring}, the Earth’s gravity gradient \cite{mcguirk2002sensitive,snadden1998measurement,sorrentino2014sensitivity,duan2014operating,fpdphysrevA,wang2016extracting,d2017canceling} and rotations \cite{gustavson2000rotation,gustavson1997precision,canuel2006six,gauguet2009characterization}. Atom interferometry experiments have been performed to test the Einstein’s Equivalence Principle \cite{tarallo2014test,rosi2017quantum,asenbaum20}, to determine the value of fundamental constants  \cite{rosi2014precision,Parker2018,Morel2020} and to search for quantum-gravity effects \cite{amelino09}. At the same time, space missions designed to study gravitational waves, dark matter and fundamental aspects of gravity using ultracold atoms have been proposed \cite{tino2019sage,elNeaj20}.

The sensitivity of light-pulse atom interferometers improves with the momentum transferred to the atoms by the beam splitters and the mirrors. Different large-momentum-transfer schemes have been demonstrated so far. They are based on repeated stimulated Raman transitions \cite{mcguirk00}, Raman composite pulses \cite{butts13}, Raman double diffraction \cite{leveque09}, Raman adiabatic rapid passage \cite{jaffe18,kotru15}, Bragg diffraction \cite{muller2008atom, kovachy15} and Floquet atom optics \cite{Wilkason2022}.
In a rubidium atom interferometer, the two hyperfine levels of the ground state provide a very good approximation of a two-level system when coupled with Raman lasers having a large detuning from the single photon transition. In the case of Bragg transitions between two momentum states, the larger is the photon recoil, the better is the two-level atom approximation. Being able to transfer high momentum in Bragg interferometry is therefore crucial both to control systematic effects and to increase the efficiency of large-momentum-transfer methods.
%In order to maintain high contrast, multi-photon Bragg diffraction processes require extremely cold atomic sources, attainable e.g. by delta-kick cooling (\textcolor{blue}{refs}), velocity selection or evaporative cooling. On the other hand, stimulated Raman transitions are widely used with laser-cooled atomic samples followed by velocity selection, thus allowing for a fast repetition of the measurement cycle. Precise atom interferometry measurements using velocity-selected sources are affected by two major decoherence mechanisms limiting the number of photon momenta transferred to the atoms: the transverse thermal motion of the atoms, which reduces the efficiency of the beam splitter and mirror pulses, and the emergence of parasitic interference paths, which increase with the number of interrogation pulses in the interferometer. 

%\todo[inline]{GMT:modifiche al par successivo}
In this paper, we report on a novel approach 
%using blue Raman lasers on the 
based on the use of Raman and Bragg pulses with the blue $5\mathrm{S}-6\mathrm{P}$ transitions of rubidium  to realize an atomic gravity gradiometer. Thanks to the higher photon recoil ($\lambda\simeq 420$~nm) compared to the  commonly used D2 line ($\lambda\simeq 780$~nm), the momentum-space splitting is increased by a factor $\sim 1.9$ while preserving the robustness and simplicity of the three-pulse Mach-Zehnder sequence. 
The shorter laser wavelength also translates into a smaller diffraction, thus reducing the systematic effects produced by the Gouy phase and wavefront distortions \cite{Wicht05,Bade18}. This scheme can find interesting applications in high precision  measurements, especially for the determination of the Newtonian gravitational constant \cite{rosi2014precision,rosi2017proposed}.

%\todo[inline]{GMT:piccole modifiche al par successivo}
We demonstrate the three-pulse Raman gravity gradiometer both on the $5\mathrm{S}_{1/2} - 6\mathrm{P}_{1/2}$ and on the $5\mathrm{S}_{1/2} - 6\mathrm{P}_{3/2}$ transitions of $^{87}$Rb at 421.7~nm and 420.3~nm, respectively \cite{glaser20}; the Bragg interferometer is operated on the $5\mathrm{S}_{1/2} - 6\mathrm{P}_{3/2}$ transition.
A critical aspect of this scheme is the laser power required to achieve high pulse efficiency and low single-photon scattering rates. For a given ratio $R$ between the two-photon Rabi frequency and the single-photon scattering rate, the Rabi frequency  is proportional to $\lambda^3$ and to the branching ratio of the transition \cite{supp}. Therefore, the smaller wavelength of the blue transition leads to a reduction of the  Rabi frequency by a factor 0.16 compared to the usual Raman interferometers operated at 780 nm. The branching ratios are 0.19 and 0.24 for the 421.7 nm and 420.3 nm transitions respectively (for relevant energy levels and decay rates see supplemental material \cite{supp}).
As a result, a higher power by a factor $\sim 30$ is required for the Raman lasers to achieve the same ratio $R$ and the same Rabi frequency as for the D2 line at 780 nm. A similar reasoning applies to the Bragg lasers.
%As a result, the optical power of the interferometry laser should be increased by a factor of roughly 30 in order to operate with the same ratio $R$ and the same Rabi frequency as in the D2 line at 780 nm. 
This technical difficulty has probably hindered so far the use of the $5\mathrm{S} - 6\mathrm{P}$ transitions for atom interferometry. For this experiment, we have developed a compact, narrow-linewidth ($\simeq$~250 kHz) blue laser system  with an output power of more than 1~W resulting in  $\sim 0.5$~W for the Raman and Bragg pulses on the atoms. It should  be noticed that two-photon ionization processes are negligible for the laser intensity and detuning from resonance employed in this work \cite{Anderlini04,supp}.

\begin{figure}[t]
\centering
%\begin{tabular}{cc}
\includegraphics[width=8.5cm]{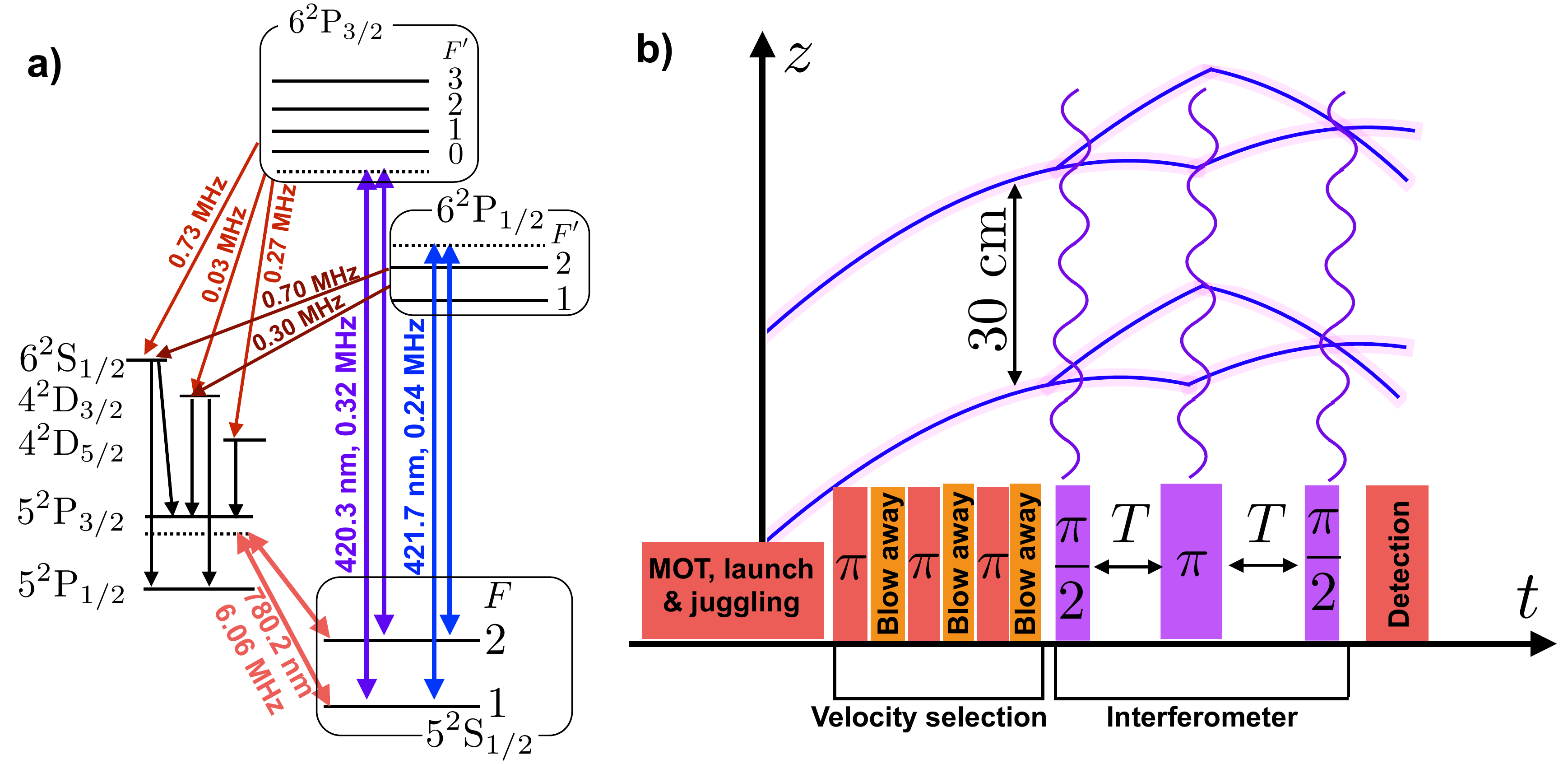}
%\end{tabular}
\caption {Experiment cycle consisting of  magneto-optical trapping, atom launch and juggling, final velocity selection and atom interferometry sequence. The velocity selection  consists of three Raman $\pi$ pulses at 780 nm alternated with blow away pulses. The two Mach-Zehnder atom interferometers are vertically separated  by 30 cm.
%  and have a duration  $2T = 320$ ms. They can be operated with Raman pulses on the D2 transition or on the $5\mathrm{S}-6\mathrm{P}$ blue transitions (violet)
%The atoms detection pulses at the end of the interferometer sequence are also shown.
}
\label{fig:fig1}
\end{figure}

%\todo[inline]{GMT:piccole modifiche al par successivo}
Our gravity gradiometer using laser-cooled Rb atoms was described in detail in \cite{rosi2014precision,prevedelli2014measuring}.
Two freely falling atomic samples separated by about 30 cm are velocity selected and prepared in the $|F=1, m_F=0\rangle$ state with a combination of Raman $\pi$ pulses and resonant blow-away laser pulses (Fig.~\ref{fig:fig1}). Infrared Raman lasers, superimposed to the blue interferometer lasers on a dichroic mirror, are used in this phase to reduce the atom losses due to off-resonance scattering. 
After atoms preparation, the two clouds, each composed of about $10^5$ atoms, are simultaneously interrogated with a $\pi/2-\pi-\pi/2$ pulse sequence by the blue Raman or Bragg lasers (Fig.~\ref{fig:fig1}). Finally, the normalized population at the two output ports of the interferometers is measured via fluorescence detection.

The blue radiation used during the interferometer sequence is obtained from two frequency-doubled infrared lasers operated in master-slave configuration.
Each of them consists of an extended cavity diode laser (ECDL) injecting a semiconductor tapered amplifier (TA) coupled to a bow-tie enhancement cavity built around a lithium triborate doubling crystal.
The master laser is stabilized on the rubidium  spectroscopy signal while the slave laser is locked on the master by a two-stage optical phase locked loop (OPLL). The primary OPLL detects the beatnote between the ECDL infrared beams before they inject the TAs. In this way, we minimize the signal propagation delay and maximize the loop bandwidth. The secondary OPLL operates on the frequency doubled beams to reduce the noise introduced by the second-harmonic generation process. The beatnote of the blue radiation from the Raman  lasers is detected before they enter the optical fiber that delivers the light to the atoms. The obtained error signal is then used to control a piezo actuator, which translates one of the mirrors along the optical path of the master Raman laser. 
Our setup is very versatile and it can be easily rearranged to generate both Raman and Bragg lasers for atoms interrogation. 

\begin{figure}[t]
%\begin{figure*}[tt]
\centering
%\begin{tabular}{cc}
%\includegraphics[width=8cm]{phaselock.pdf}&
%\includegraphics[width=8cm]{phasenoise.pdf}\\
\includegraphics[width=8.5cm]{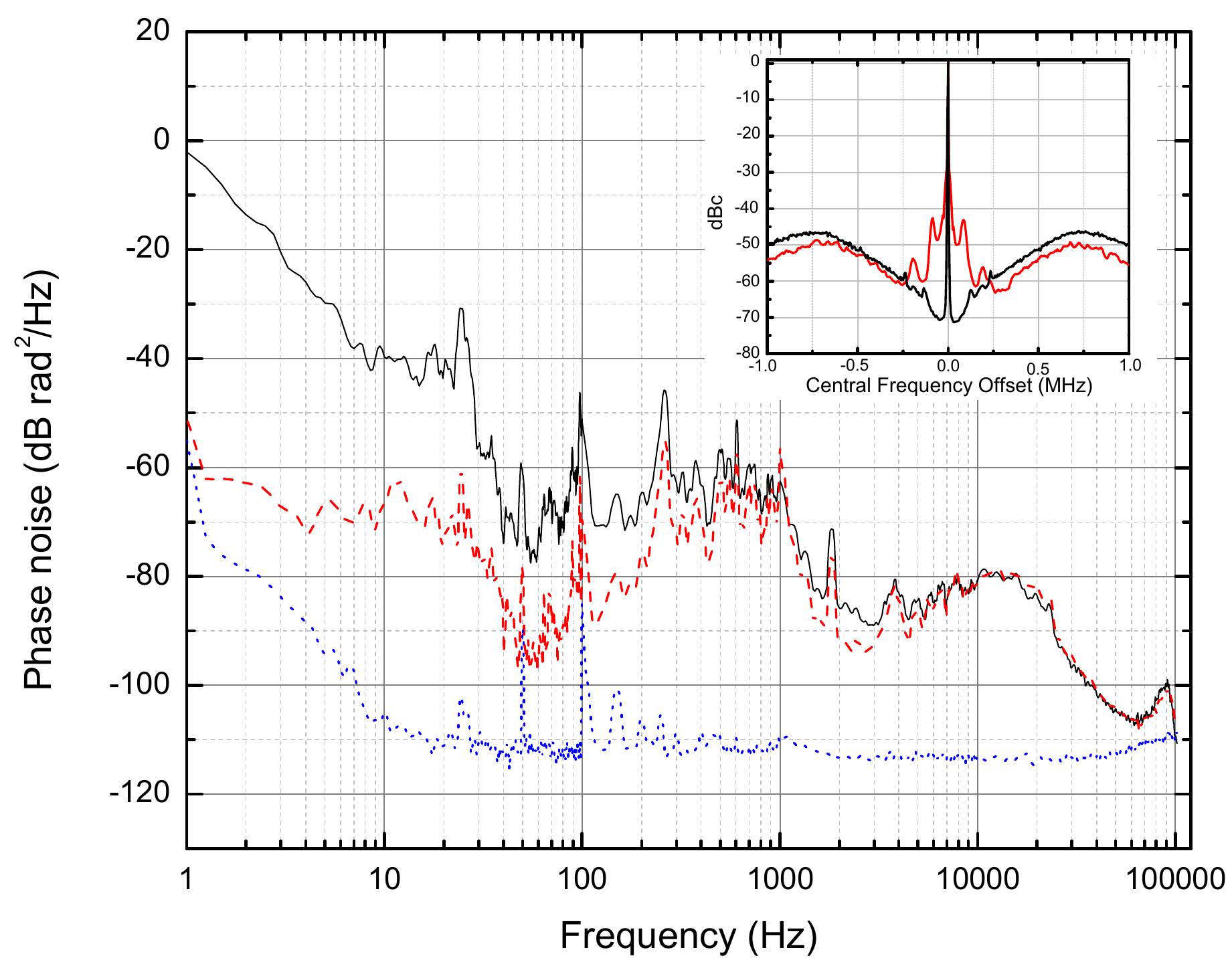}
%\end{tabular}
\caption{Phase noise power spectra of the infrared beatnote with the primary OPLL closed (dotted blue line) and of the blue beatnote with the secondary OPLL open (black line) and closed (dashed red line). Inset: Beatnote signal between the two Raman lasers on the fundamental (black line) and frequency doubled (red line) light with the primary OPLL closed (resolution bandwidth: 10 kHz). %\textcolor{red}{distinguere le curve con numerazione e trattini, punti, ecc.}
}
\label{fig:fig6}
%\end{figure*}
\end{figure}

To evaluate the contribution of the frequency doubling process to the interferometry lasers phase noise, we characterize the master-slave beatnote both at the fundamental frequency and at the second harmonic.
Figure \ref{fig:fig6} (inset) shows the beatnote signals  in the infrared (black line) and in the blue (red line) after closing the primary OPLL. The frequency doubling process increases the phase noise at frequencies below 250~kHz. On the contrary, a reduction of the blue beatnote servo-bumps is observed, due to the filtering effect introduced by the finite width ($\simeq500$~kHz HWHM) of the doubling cavity resonance. Figure~\ref{fig:fig6} shows the phase noise power spectrum of the infrared beatnote when the primary OPLL is closed (blue line) and of the blue beatnote with the secondary OPLL open (black line) and closed (red line). The excess of noise introduced by the frequency doubling process increases at an average rate of 20~dB/decade as the frequency decreases. The secondary OPLL reduces the noise below 1~kHz and restores the phase coherence. This is sufficient for our experiment due to the excellent suppression factor that a gravity gradiometer can ensure on the common-mode phase noise of the interrogation lasers. For the operation of the interferometer as an accelerometer, this performance, currently limited by the bandwidth of our piezo, can be improved and extended to higher frequencies by using a faster actuator (e.g. an electro-optical modulator).

We characterize the Raman gravity gradiometer sensitivity on both the $5\mathrm{S}_{1/2} - 6\mathrm{P}_{1/2}$ and $5\mathrm{S}_{1/2} - 6\mathrm{P}_{3/2}$ transitions and compare it to the interferometer on the D2 line. The Raman beams are collimated to a waist of $9$~mm for the blue and $13$~mm for the infrared. With respect to the typical parameters of our infrared Raman interferometer, we maintain a pulse separation of $T=$ 160~ms and a repetition rate of 0.5~Hz, but we increase the $\pi$ pulse duration from 24 to 48~$\mu$s. The intensity ratio $R_{21}$ between the lasers interacting with the $F=2$ and $F=1$ levels of the ground state is tuned to cancel the differential AC Stark shift \cite{peters2001high, supp}. For the interferometer on the $5\mathrm{S}_{1/2} - 6\mathrm{P}_{1/2}$ transition, we use a total maximum Raman power of 0.44 W (measured after the fiber), an intensity ratio $R_{21}\approx 1$, and a positive detuning of 50 MHz from the $F=2-F'=2$ transition. On the $5\mathrm{S}_{1/2} - 6\mathrm{P}_{3/2}$ transition,  we have a total maximum power of 0.55 W and $R_{21}\approx 2$, with a negative detuning of 80 MHz from the $F=2-F'=1$ transition. The spontaneous emission probability for the overall interferometer is $\sim 13$\%  for both blue wavelengths. 
%During the interferometer sequence, Raman lasers are frequency chirped in opposite directions so that their frequencies are fixed in the atomic frame. This is important for maintaining a constant pulse efficiency during the interferometer since we are dealing with small transition detuning.
\begin{figure}[t]
   \centering
\begin{tabular}{cc}
\includegraphics[width=8.3cm]{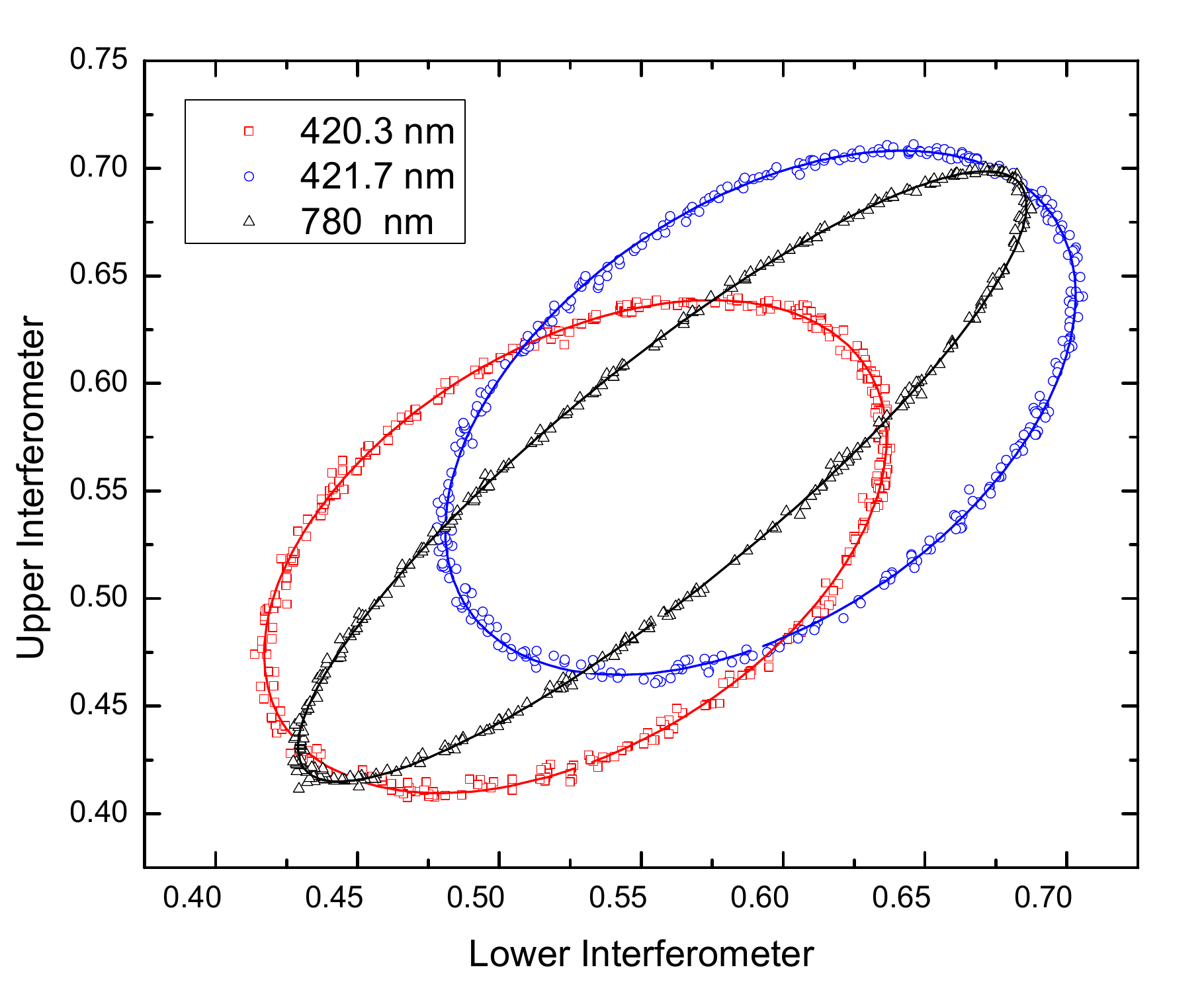}
\end{tabular}
\caption{Experimental ellipses (360 points each) obtained with Raman interferometry at 420.3 nm (red open squares), at 421.7 (blue open circles) and at 780 nm (black triangles). Elliptical fits are indicated with red, blue and black solid lines, respectively. To avoid overlaps and improve clarity, the centers of the ellipses in the figure have been translated.}
\label{fig:fig7}
\end{figure}
Figure \ref{fig:fig7} shows typical Lissajous curves obtained by plotting the interference fringes of the upper interferometer as a function of the fringes of the lower interferometer when using Raman pulses on the blue transitions (420.3 and 421.7 nm) and on the infrared transition (780 nm) during the interferometer sequence. By comparing the results obtained at the three different wavelengths, we experimentally confirm the expected increase by a factor 1.9 of the gradiometer phase shift: from the 0.6~rad observed at 780~nm to the 1.1~rad obtained with the blue transitions. Figure \ref{fig:fig7} also shows that the fringe contrast is reduced by 10-20\% with respect to the interferometer on the infrared transition, which is partially explained by spontaneous emission. Moreover, the ellipses obtained on the blue transitions reveal a higher noise level. After analysing the atomic fluorescence signals at detection, we could attribute this extra noise to the higher single-photon scattering rate on the blue transitions, which affects the atom counting at the output ports of the interferometers \cite{supp}.
%We attribute this noise to the relatively high single-photon scattering rate in our measurement conditions. It is indeed manifest in the atomic fluorescence signals that exhibit a significant thermal background produced by the blue laser beam pulses.
To characterize the instrument sensitivity and the medium-term stability, we evaluate the Allan deviation of the gravity gradient measurements normalized to the local gravitational acceleration
%fractional differential gravity sensitivity
over a total measurement duration of about 6~h. As shown in Fig.~\ref{fig:fig8}, the Allan deviation decreases as $1/\sqrt{t}$, where $t$ is the integration time, showing that gravity gradient measurements are affected by a white noise process. In this respect the interferometers operated on the blue and infrared transitions have similar performance within a maximum difference of 30\%. 
Indeed, for our experimental conditions, the higher signal ($\times 1.9$) measured on the blue transitions is compensated by the higher noise levels that we observe due to single-photon scattering processes. This important noise contribution can be reduced considerably by using higher power laser sources that allow to operate the Raman lasers at a larger detuning from the excited states. The resulting interferometer sensitivity to differential acceleration measurements is $1\times10^{-8}$ g at 1~s, reaching $2\times10^{-10}$g after 2000 s of integration time. 

\begin{figure}[t]
   \centering
\begin{tabular}{cc}
\includegraphics[width=8.5cm]{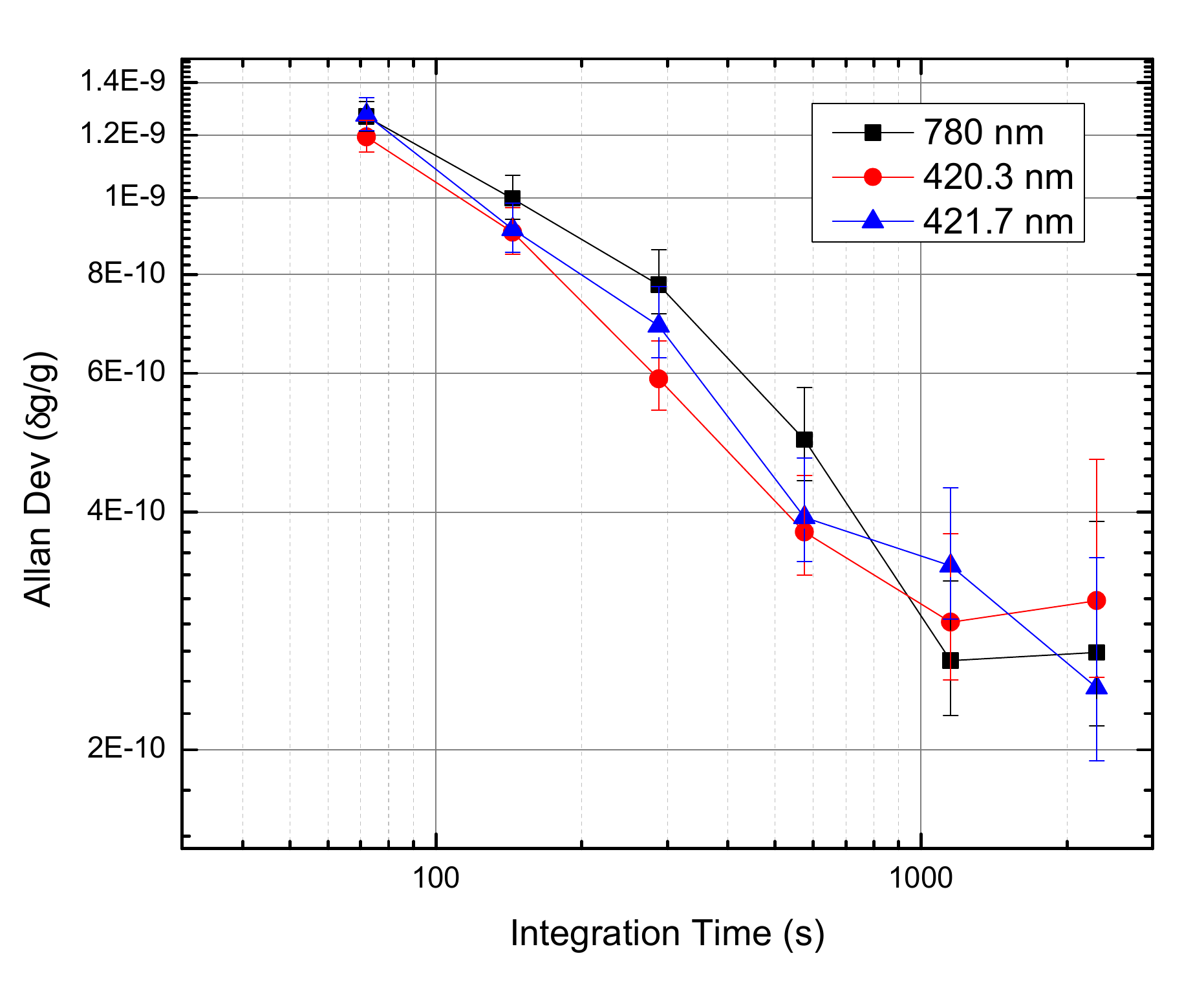}
\end{tabular}
\caption{Allan deviation plots of the gravity gradiometer measurements normalized to the local gravitational acceleration, obtained with Raman interferometry at 420.3 nm (red diamonds), at 421.7 nm (blue triangles) and at 780 nm (black squares). %\textcolor{red}{should we put this in units of differential acceleration as cited in the text? Double vertical axis.}
}
\label{fig:fig8}
\end{figure}

\begin{figure}[t]
   \centering
\begin{tabular}{cc}
\includegraphics[width=8.5cm]{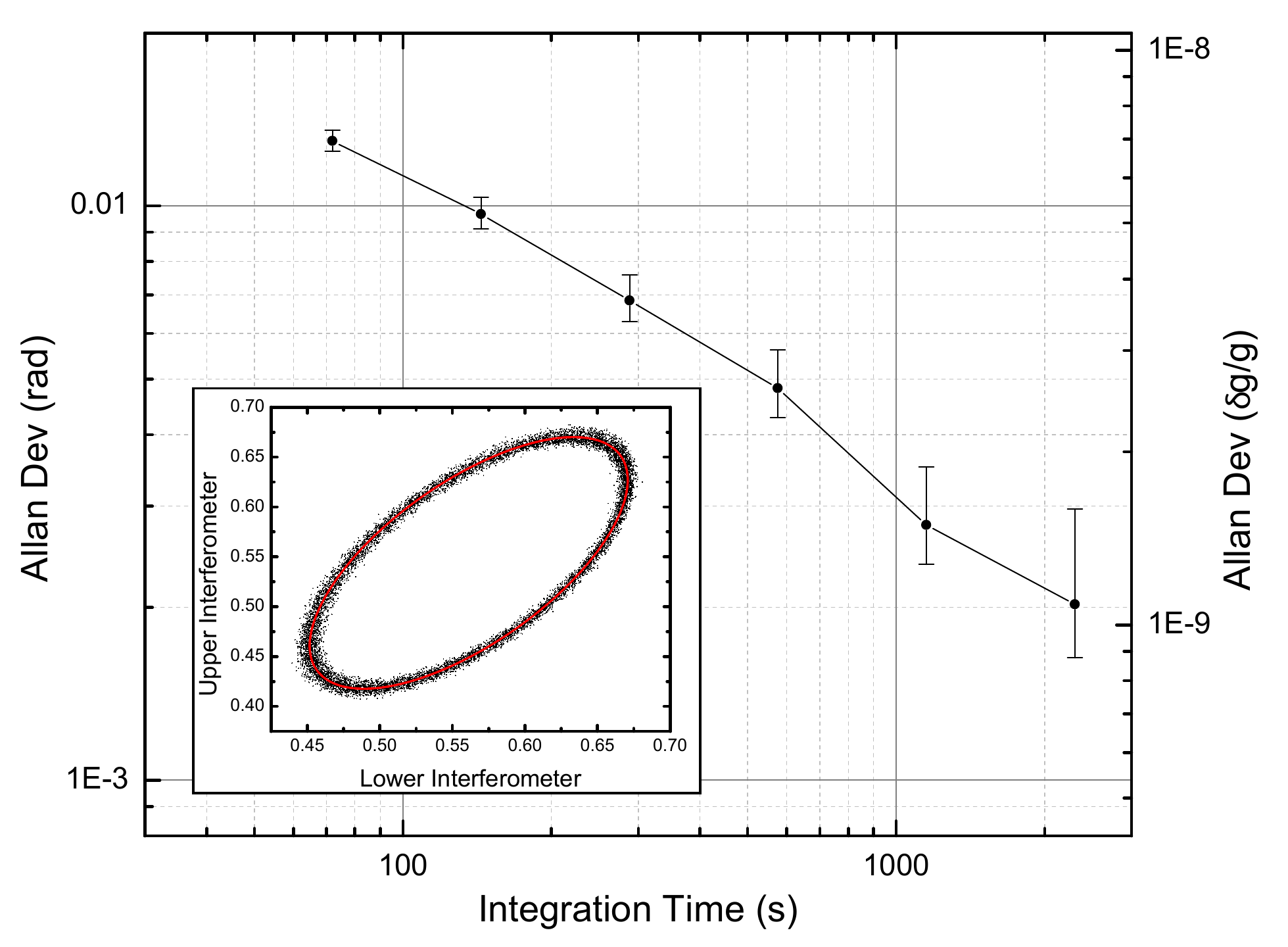}
\end{tabular}
\caption{Allan deviation plot of the gravity gradiometer measurements normalized to the local gravitational acceleration obtained with Bragg interferometry at 420.3~nm. Inset: Lissajous plot of the 12500 data points collected during the measurement run, together with the elliptical best fit (red line). %\textcolor{red}{should we put this in units of differential acceleration as cited in the text? Double vertical axis.}
}
\label{fig:fig9}
\end{figure}

In a second experiment, we tune the blue lasers on the $5\mathrm{S}_{1/2} -  6\mathrm{P}_{3/2}$ transition at 420.3~nm and configure them to operate the gravity gradiometer with first-order Bragg pulses. 
To reduce the single-photon scattering rate and consequently the noise on the atom number measurements at the output ports of the interferometer, Bragg lasers are set to a negative detuning of 252~MHz with respect to the $F=1 - F'=1$ transition, about three times larger than for the Raman interferometer. As a consequence, for a total laser power of 0.5~W, the beam waist had to be reduced to 7~mm. Rubidium atoms are interrogated in the Bragg interferometer on a $\pi/2-\pi-\pi/2$ sequence of Gaussian pulses. The $\pi$ pulse has a Gaussian standard deviation of 21~$\mu$s and a separation of $T=80$~ms with respect to the beam splitter pulses. Pulse duration is chosen as a compromise between reducing losses to undesired diffraction orders and optimizing the velocity acceptance. The three-pulse sequence takes place during the ascending phase of the atomic clouds to maximize the spatial separation between the two output ports of the interferometers. Due to the higher photon recoil carried by the blue lasers, first-order Bragg transitions are already sufficient to use the standard position-resolved detection systems for atom counting. Momentum-resolved detection methods \cite{Cheng18} will also benefit from the larger velocity difference. Finally, during the interferometric sequence we change the frequency detuning of the Bragg $\pi$ pulse by 40~MHz to introduce a differential phase shift and mimic the presence of a gravity gradient \cite{roura2017,DAmico17}. In this way, we can further open the ellipse and improve the best fit parameter estimates. 
Figure~\ref{fig:fig9} (inset) shows the Lissajous plot obtained from the gravity gradiometer measurements using Bragg-pulse interferometry. The fringes contrast is similar to the one observed for the Raman interferometers on the blue transitions, but it is obtained at a significantly larger frequency detuning and for a shorter interferometer duration. As for the Raman interferometer, spontaneous emission is the main contributor to the noise that we observe in our measurements.
The stability of the first-order Bragg gravity gradiometer is characterized by evaluating the Allan deviation of the differential acceleration measurements (see Fig.~\ref{fig:fig9}). The slope of the Allan deviation curve indicates the presence of a white noise process. The instrument stability is $5.9\times10^{-8}$~g at 1~s, averaging down to $1\times10^{-9}$~g after 2000~s of integration time. Compared to the results obtained with the third-order Bragg gravity gradiometer on the infrared transition characterized in \cite{DAmico16}, the sensitivity to differential acceleration measurements is degraded by a factor 1.6, which exactly corresponds to the ratio of the photon momenta transferred to the atoms in the two experiments. This result would confirm that the factor 1.9 improvement on the instrument sensitivity can indeed be obtained when the Bragg lasers have adequate detuning from the single-photon transition. The laser power available in the blue is not sufficient to operate our gravity gradiometer on higher-order Bragg transitions therefore a direct comparison, blue vs infrared, is not possible.

In conclusion, we have demonstrated Raman-pulse and Bragg-pulse atom interferometry on the 420.3~nm and 421.7~nm transitions of $^{87}$Rb.  The larger momentum transferred to the atoms on the blue transitions provides an increase by a factor 1.9 of the gravity gradiometer differential phase with respect to the infrared transition at 780 nm, while preserving the simplicity and robustness of the three-pulse Mach-Zehnder interferometer. With a stability of $1\times10^{-8}$ g at 1~s, the Raman gravity gradiometer noise averages down to $2\times10^{-10}$g after 2000~s of integration time at both blue wavelengths. When the atoms are interrogated in the first-order Bragg interferometer, we obtain a stability of $5.9\times10^{-8}$~g at 1~s, reaching $1\times10^{-9}$~g after 2000~s of integration time.
%Further improvements are possible.
%to surpass the one obtained in the infrared with the same apparatus \cite{sorrentino2014sensitivity}. 
%
%In this work, the blue laser light  was generated by a home-made laser system delivering up to 1.4~W output power.
%with narrow linewidth (255~kHz) and low relative intensity noise (-100~dBc).
%
Due to the higher intensity required for the blue transitions compared to the infrared, the blue interrogation lasers had to be shaped to a smaller beam diameter and operated to deliver longer $\pi$ pulses at a relatively smaller detuning from the resonance with the single-photon transition. In this configuration, the lower number of atoms at detection and the higher spontaneous emission rate reduce the fringe contrast and introduce substantial noise in the differential phase measurement. 
As also shown in the Bragg-pulse interferometry experiment, higher laser power and larger detuning from resonance will be required to reach optimal working conditions. The laser power can be increased by adding an amplification stage on the infrared radiation before the second-harmonic generation or by injection-locking schemes of Ti:sapphire lasers that have recently been demonstrated to produce several watts of infrared light with an excellent spatial profile \cite{takano21}. We expect that such an upgrade will allow to reduce the noise due to spontaneous emission and unravel the full potential of the presented interferometer scheme. We also point out that the near-unity intensity ratio of the Raman beams for which the differential AC Stark shift is canceled at 421.7 nm allows to optimally use the available laser power as the two-photon Rabi frequency is maximized by this ratio. Moreover, the deviation from the unity ratio remains on the order of 10\% for a positive detuning up to 300 MHz. For the 420.3 nm transition, the optimal intensity ratio is found at about 1.36 and it is detuning-independent to first order for a significantly higher negative detuning of 458 MHz \cite{supp}. This is important when implementing interrogation schemes requiring a change of the laser frequency to mimic the presence of a gravity gradient \cite{roura2017,d2017canceling}, as also done in this work.  
Large-momentum-transfer schemes already demonstrated with the infrared transition can be implemented in the blue to further increase the gradiometer signal. 
%The improved phase-space densities that can be achieved in a blue $^{87}$Rb MOT \cite{Jarvis18} can also be used to increase the atom number contributing to the interferometer signal. 
With this respect, the larger photon recoil transferred by the blue lasers in a Bragg transition is also providing a larger separation between the two coupled momentum states thus promising a better control of systematic effects and a higher efficiency of the large-momentum-transfer pulses.
Furthermore, the shorter laser wavelength will lead to a reduction of systematic effects like the Gouy phase and wavefront distortion currently limiting several atom interferometry experiments.
Therefore, we anticipate that the scheme demonstrated in this work will be instrumental for high-precision experiments as, in particular, for  the determination of the Newtonian gravitational constant \cite{jain21}.

\begin{acknowledgments}
This work was supported by INFN under the OLAGS project and the QuantERA grant SQUEIS. We acknowledge financial support from: PNRR MUR Project No. PE0000023-NQSTI.
GR acknowledges financial support from the European Research Council, Grant No. 804815 (MEGANTE) and from MIUR (Italian Ministry of Education, Universities and Research) under the FARE-TENMA project. 
LS acknowledges support from Università di Firenze under the QuEGI-Quantum Enhanced Gravity Interferometry project. 
The authors thank Giovanni Gangale for help in an early stage of the experiment.

\end{acknowledgments}

\bibliography{aipsamp.bib}

\end{document}